\documentclass[a4paper,fleqn]{cas-dc}

\usepackage[sort&compress,numbers]{natbib}

\usepackage{xcolor}

\begin{document}
\let\WriteBookmarks\relax
\def\floatpagepagefraction{1}
\def\textpagefraction{.001}

\shortauthors{A. Alekseev, A. Cichy and K. J. Kapcia}
%
%
\title [mode = title]{Charge order and stripes in the $t$-$V$ model on the kagome lattice at the ground state}

\shorttitle{Charge order and stripes in the $t$-$V$ model on the kagome lattice at the ground state}

\author[1]{Aleksey Alekseev}[%
                    orcid=0000-0001-5102-6647]
\ead{aleksey.alekseev@amu.edu.pl}
\credit{Conceptualization, Methodology, Software, Formal analysis, Investigation, Visualization, Writing - original draft, Writing - review and editing}

\author[1,2]{Agnieszka Cichy}[%
                    orcid=0000-0001-5835-9807]
\ead{agnieszka.kujawa@amu.edu.pl}
\credit{Formal analysis, Supervision, Validation, Writing - original draft, Writing - review and editing}

\author[1]{Konrad Jerzy Kapcia}[%
                    orcid=0000-0001-8842-1886]
\ead{konrad.kapcia@amu.edu.pl}
\credit{Conceptualization, Methodology, Data curation, Formal analysis, Investigation, Funding acquisition, Project administration, Resources, Supervision, Validation, Writing - original draft, Writing - review and editing}

\address[1]{Institute of Spintronics and Quantum Information, Faculty of Physics and Astronomy, Adam Mickiewicz University in Pozna\'n, Uniwersytetu Pozna\'nskiego 2, PL-61614 Pozna\'n, Poland}
\address[2]{Institut f\"ur Physik, Johannes Gutenberg-Universit\"at Mainz, Staudingerweg 9, D-55099 Mainz, Germany}

\begin{abstract}
The kagome lattice can host  {unexpected} or anomalous phenomena, but the whole picture of what can be realized in materials with the kagome structure is difficult to obtain. 
Here, as a step toward such  {knowledge}, we investigate the kagome-lattice spin-$1/2$-particle $t$-$V$ model (i.e., onsite interactions are absent, only nearest-neighbor density-density interactions are included) without fixing the total charge density,  {and solve it using} the mean-field approximation (MFA). 
The MFA phase diagram contains  {many} interesting features, such as unexpected phases, varying order of phase transitions (tricritical points), strong particle-hole asymmetry, and quasi-1D structures with charge itineracy in one direction only. 
We expect that these features can also appear when the correlation effects are taken into account (e.g., due to the Hubbard onsite interaction), which would result in  {an} even more complex phase diagram that would be better interpreted with  {access} to comprehensive MFA  {studies}. 
Note that since the MFA disregards the local correlations and does not track rigorously nonlocal correlations, it  {allows} band-structure analysis, which provides clarity about the phenomena that can still be found with more advanced methods.
\end{abstract}

%
%
%
%
%
%
%
\maketitle
%
%
%

\section{Introduction}

Materials with  {a} kagome lattice  {have} received  {considerable} of attention  {because of their} unique properties,  {which give rise to states such} as superconducting phases, exotic ordered phases, and more.
In this work we focus on a charge-ordering phenomenon on the kagome lattice, which, besides being interesting on its own,  {interplays with} or competes with superconductivity \cite{denner2021analysis, christensen2021theory, mielke2022time, neupert2022charge, ferrari2022charge, kang2022twofold, dong2023loop, khasanov2022time,  guguchia2023tunable, jiang2021unconventional, shumiya2021intrinsic, guo2022switchable, PtokPRB2022,  PtokPRB2021, zhang2024electronic, lee2024nature, cheng2025broken, tuniz2025strain, holbaek2025interplay, SajidPRB2026}. 

We investigate a kagome-lattice spin-$1/2$ $t$-$V$ model  {with three-sublattice assumption} from the perspective of  {the} \mbox{$U \rightarrow 0$} limit of the extended Hubbard model, where $U$ is an onsite interaction strength. 
We solve it within the mean-field approximation (MFA) which is advantageous  {because it is} computationally fast and  {allows} to analyze and interpret results using band structures. 
The essence of the MFA is to disregard all correlations, thus neglecting a large number of most of the interesting strongly correlated phenomena. 
Having said that, the peculiarity of the kagome lattice results in a quite complex MFA phase diagram, full of various features that can still be found after the correlations are taken into account. 
The $U \neq 0$ case should be  {treated} within the dynamical mean-field theory (DMFT) \cite{GeorgesRMP1996, ImadaRMP1998}, 
but the MFA results already contain a variety of interesting and unexpected features that could be found within the DMFT and  {can be interpreted much more clearly} in the MFA, 
as it was found for a triangular lattice that also hosts a frustrated geometry \cite{alekseevPRB2026, AlekseevPRB2025}. 
Additionally, the presented results can further be viewed as  {a starting point at $U=0$} of the DMFT results of the extended Hubbard model. 
They can help to identify what features come from strong correlations or from the finite onsite interaction. 
In our analysis we also utilize an atomic-limit prediction for the phase diagram. 

A distinctive aspect of our research is arbitrariness of the total charge density (filling factor).
We work in the grand canonical ensemble and use chemical potential as a variable, thus avoiding restriction of the total charge density,  {thereby allowing us to investigate the entire} phase diagram \cite{alekseevPRB2026, AlekseevPRB2025, LitakJMMM2017}, and naturally taking into account a phase-separation phenomenon \cite{BakAPPA2004}. 
Note that even the common models are  {insufficiently investigated when the total charge density is not fixed} and various physical properties are potentially missed. 
In particular, most existing theoretical studies of charge order on the kagome lattice focus on specific materials, particularly the AV$_3$Sb$_5$ family, and therefore work at or near a fixed, physically motivated filling (typically the Van Hove filling at $n = 5/6$ per lattice site), e.g.,  \cite{khasanov2022time, guguchia2023tunable, jiang2021unconventional, shumiya2021intrinsic, guo2022switchable, PtokPRB2022}.
Ultimately, even the MFA results, as a first step before the DMFT and its extensions, 
reveal many interesting features of ground-state phase diagram for the model studied, which require detailed discussion.

Advantageously, we do not limit the research to any particular kagome-lattice material, which helps to find new potentially interesting phases and  {obtain a broader} picture and physical mechanisms behind it. 
It is especially relevant since the models can now be reproduced experimentally with ultra-cold atomic gases on optical lattices, quantum simulations, or even moir\'e lattices \cite{GeorgescuRMP2014, DuttaRPP2015, Wu2018}.

The nonlocal correlations are known to be very important for 2D lattices being much amplified by the geometrical frustration. 
We expect, that the extended DMFT, that takes nonlocal correlations into account, should benefit from our MFA investigation the same way as the it helps to interpret the non-extended (i.e., single-site) DMFT results with $U \neq 0$ (for the triangular lattice see \cite{AlekseevPRB2025, alekseevPRB2026} whereas for the Bethe lattice see \cite{Kapcia2017, AlekseevPhysicaA2026}).

\section{Model and method}\label{sec:method}

The Hamiltonian of the spin-$1/2$ $t$-$V$ model considered in this work has the following form
\begin{equation}\label{eq:model}
    \hat{H} 
    = - t \sum_{\langle i,j \rangle, \sigma} \left( \hat{c}^\dag_{i\sigma} \hat{c}_{j\sigma} + \text{H.c.} \right)
    + V \sum_{\langle i,j \rangle} \hat{n}_i \hat{n}_j
    - \mu\hat{N},
\end{equation}
where $t$, $V$, and $\mu$ are a hopping amplitude, intersite nearest-neighbor (NN) density-density interaction strength, and chemical potential of spin-$1/2$ particles, respectively;
$i$ and $j$ are lattice-site indices, 
$\sigma=\uparrow,\downarrow$ is a spin index, 
$\sum_{\langle i,j \rangle}$ means summation over NN pairs; 
$\hat{c}^\dag_{i\sigma}$ and $\hat{c}_{i\sigma}$ are creation and annihilation fermion operators (with standard anticommuting relations), 
$\hat{N} = \sum_i \hat{n}_i$,
$\hat{n}_i = \sum_\sigma \hat{n}_{i\sigma}$, 
and $\hat{n}_{i\sigma} = \hat{c}^\dag_{i\sigma} \hat{c}_{i\sigma}$.

The Hamiltonian (\ref{eq:model}) does not have explicit spin-dependent interactions but includes a spin-$1/2$ degeneracy primarily because we are interested in the extension of our research to more general cases.
With the Hubbard term $U\sum_i \hat{n}_{i\uparrow} \hat{n}_{i\downarrow}$ (cf. \cite{HubbardPRSL1963, PennPR1966, Spalek1987}) the model (\ref{eq:model}) becomes the extended Hubbard model \cite{MicnasRMP1990, PietigPRL1999, TongPRB2004, AmaricciPRB2010, Kapcia2017, AlekseevPRB2025, AlekseevPhysicaA2026, alekseevPRB2026}, where $U$ is the onsite interaction strength,  discussed occasionally throughout this paper  (cf. also Ref.  \cite{KapciaJMMM2020}, where model (\ref{eq:model}) with the spin-asymmetric hopping was investigated).

\subsection{Mean-field approximation}

The general methodology for the mean-field calculations is the same as  {described} in Ref. \cite{AlekseevPRB2025}. After
\begin{itemize}
    \item[(i)] a Fourier transform
    from a lattice site coordinate $\mathbf{R}_i$ to a reciprocal vector $\mathbf{k}$ defined as
    $\hat{c}_{\alpha\mathbf{k}\sigma} = \sqrt{(3/L)} \sum_{i\in\alpha} \hat{c}_{i \sigma} e^{- \mathbb{i} \mathbf{k} \cdot \mathbf{R}_i}$,
    where $L$ is a number lattice sites enclosed by periodic boundary conditions, and the dot product is defined as $\mathbf{k} \cdot \mathbf{R}_i = 2\pi (k_x R_{ix} + k_x R_{iy})$;
    \item[(ii)] an assumption of periodicity of a kagome-lattice primitive cell, meaning  that all found orders are commensurate with the cell with three lattice sites (sublattice index $\alpha = 1, 2, 3$);
    \item[(iii)] and the mean-field (Hartree) decoupling 
    $\hat{n}_{i\sigma} \hat{n}_{j\sigma'} \overset{\text{MF}}{=} n_{i\sigma} \hat{n}_{j\sigma'} + n_{j\sigma'} \hat{n}_{i\sigma} - n_{i\sigma} n_{j\sigma'}$,
\end{itemize}
the Hamiltonian (\ref{eq:model}) takes the form
\begin{equation}\label{eq:MFmodel}
    \hat{H} 
    \overset{\text{MF}}{=} \sum_{\mathbf{k}\sigma} 
    \hat\Psi^\dag_{\mathbf{k}\sigma}
    \mathbb{H}_\mathbf{k}
    \hat\Psi_{\mathbf{k}\sigma}
    + C,
\end{equation}
where 
$\hat\Psi_{\mathbf{k}\sigma} = ( \hat{c}_{1\mathbf{k}\sigma}, \hat{c}_{2\mathbf{k}\sigma}, \hat{c}_{3\mathbf{k}\sigma} )$,
$\hat{c}_{\alpha\mathbf{k}\sigma}$ is an annihilation operator of a particle that belongs to a sublattice $\alpha$ and has a momentum $\mathbf{k}$ and spin $\sigma$; 
the matrix
\begin{equation}\label{eq:Hmat}
    \mathbb{H}_\mathbf{k} =
    \begin{pmatrix}
        \epsilon_1 & \varepsilon_{12\mathbf{k}} & \varepsilon_{13\mathbf{k}} \\
        \varepsilon_{12\mathbf{k}} & \epsilon_2 & \varepsilon_{23\mathbf{k}} \\
        \varepsilon_{13\mathbf{k}} & \varepsilon_{23\mathbf{k}} & \epsilon_3
    \end{pmatrix},
\end{equation}
with elements
\begin{equation}\label{eq:energies}\begin{aligned}
    \epsilon_\alpha &= \frac{zV}{2} (n_{\bar\alpha} + n_{\bar{\bar\alpha}}) - \mu,
    \\
    \varepsilon_{12\mathbf{k}} &= -2t\cos{\pi k_x},
    \\
    \varepsilon_{13\mathbf{k}} &= -2t\cos{\pi k_y},
    \\
    \varepsilon_{23\mathbf{k}} &= -2t\cos{\pi (k_x + k_y)},
\end{aligned}\end{equation}
where
$z=4$ is a coordination number,
$\bar\alpha$ and $\bar{\bar\alpha}$ are sublattice indices that are different from $\alpha$ and from each other, 
$n_i = \langle\hat{n}_i\rangle = \sum_\sigma n_{i\sigma}$, 
$n_{i\sigma} = \langle\hat{n}_{i\sigma}\rangle$;
vector $\mathbf{k} = (k_x, k_y)$ is written in the basis of a reciprocal cell;
and the constant
\begin{equation}
    C = - \frac{L}{3} \frac{zV}{2} \sum_\alpha n_{\bar\alpha} n_{\bar{\bar\alpha}}.
\end{equation}

Further, we express most energy-related quantities in the units of $D=3t$ which is a half-bandwidth of a noninteracting kagome lattice. 
Moreover, the intersite interaction $V$ is expressed in the units of $D/z = 0.75t$ for the sake of better consistency of results for different lattices.

The eigenvalues and eigenvectors of the matrix (\ref{eq:Hmat})
are $E_{\mathbf{k}m}$
and $\mathbf{a}_{\mathbf{k}m} = \left( a_{1\mathbf{k}m}, a_{2\mathbf{k}m}, a_{3\mathbf{k}m} \right)$ with a band index $m=1,2,3$. 
With that, the occupation numbers are
\begin{equation}\begin{aligned}
    n_\alpha &= \frac{3}{L} \sum_{i\in\alpha} n_i = \frac{3}{L} \sum_{\mathbf{k}\sigma} \langle \hat{c}^\dag_{\alpha\mathbf{k}\sigma} \hat{c}_{\alpha\mathbf{k}\sigma} \rangle
    \\&= \frac{6}{L} \sum_{\mathbf{k} m} |a_{\alpha\mathbf{k}m}|^2 f_\text{FD}(E_{\mathbf{k}m}),
\end{aligned}
\end{equation}
and the model can be solved self-consistently. Here, $f_\text{FD}(\omega) = (1 + e^{\beta \omega})^{-1}$ is a Fermi-Dirac distribution and $\beta = T^{-1}$ is  the inverse temperature. This work is focused on the ground-state properties (zero temperature), however, due to convergence problems, the calculations are performed at $T=10^{-5}D$. 
On the scale of presented results they do not change for any further decrease of $T$; thereby we refer to our results as the ground-state results. 
Moreover,  an extremely strong mixing  procedure (such as  $n_{\alpha,\text{new}} = 0.995n_{\alpha,\text{old}} + 0.005n_{\alpha,\text{new}}$) is used when  the Fermi level  is in the vicinity of a singularity of a spectral function. 
Various convergence problems are the same as for Bethe \cite{AlekseevPhysicaA2026} and triangular \cite{AlekseevPRB2025} lattices including other self-consistent methods such as DMFT \cite{alekseevPRB2026}. The self-consistency convergence of our results is far beyond the presented scale, such as $10^{-9}$ to $10^{-12}$ difference in $n_\alpha$ between iterations.

We use the $96 \times 96$ and (for better final plots) $360 \times 360$ $\mathbf{k}$-point Monkhorst-Pack grids for the $\mathbf{k}$-point summation  for which the convergence of the results has been ensured. 
Usually, the Monkhorst-Pack grid is not compatible with the hexagonal-lattice symmetry \cite{monkhorst1976special}, however, the most-broken-symmetry phase, where all three sublattices are nonequivalent, does not host the hexagonal symmetry (see Sec. \ref{sec:sym-bands}), and its symmetry is used across all calculations.

The stability of the phases is determined by comparing their grand potentials (per lattice site)
\begin{equation}
    \frac{\Omega}{L} = \frac{2}{L} \sum_{\mathbf{k}m} \left( E^\text{GS}_{\mathbf{k}m} - \frac{1}{\beta}\ln \tilde{\mathcal{Z}}_{\mathbf{k}m} \right)
    + \frac{C}{L},
\end{equation}
where $E^\text{GS}_{\mathbf{k}m} = \min(0, E_{\mathbf{k}m})$, 
and 
\begin{equation}
    \tilde{\mathcal{Z}}_{\mathbf{k}m} = \mathcal{Z}_{\mathbf{k}m} e^{\beta E^\text{GS}_{\mathbf{k}m}} = e^{\beta E^\text{GS}_{\mathbf{k}m}} + e^{-\beta (E_{\mathbf{k}m} - E^\text{GS}_{\mathbf{k}m})}
\end{equation}
(it is numerically stable to count energy exclusively from $E^\text{GS}_{\mathbf{k}m}$ for large $\beta$).

The results are shown with respect to the shifted chemical potential $\bar\mu = \mu - zV$, that is chosen so that the properties of a half-filled non-ordered phase do not depend on $V$ for a fixed $\bar\mu$, even though for the kagome lattice it happens for $\bar\mu \ne 0$, and the kagome lattice does not exhibit a particle-hole symmetry. 
Moreover, $\bar\mu = 0$ is a point of the particle-hole symmetry in the atomic limit of the model (see Sec. \ref{sec:AL}).

\subsection{System properties and order parameters}

Three order parameters that partially identify the found self-consistent solutions of the model (\ref{eq:MFmodel}) are occupation numbers $n_\alpha$. 
Let us also introduce the total charge density $n = \frac{1}{3} \sum_\alpha n_\alpha$.

An advantage of the MFA is that it allows us to analyze the band structure, i.e., $\omega_m(\mathbf{k}) \equiv E_{\mathbf{k}m}$, the eigenvalues of the matrix (\ref{eq:Hmat}). 
We also examine the sublattice-specific spectral functions $A_\alpha(\omega + i\eta)$, where $\eta = 0.005D$,
\begin{equation}
    A_\alpha(z) = -\frac{2}{\pi} \frac{3}{L} \sum_\mathbf{k} \operatorname{Im} G_{\alpha\alpha\mathbf{k}}(z),
\end{equation}
$G_{\alpha\alpha\mathbf{k}}(z)$ are diagonal elements of the matrix $(zI - \mathbb{H}_\mathbf{k})^{-1}$, and $I$ is a $3 \times 3$ identity matrix.

To estimate insulating properties of a sublattice we also pay attention to three more order parameters, spectral weights at the Fermi level (i.e., at $\omega=0$), $A_\alpha(i\eta)$. 
When the whole system (all three sublattices) is an insulator, it can also be seen from the Fermi level located in a band gap. 
Moreover, insulating phases always have strictly integer values of $3n = \sum_\alpha n_\alpha$ (number of particles per unit cell).

Note that $A_\alpha(\omega + i\eta)$ is very sensitive to the choice of $\eta$, and the exact value of $A_\alpha(\omega + i\eta)$ is never used to draw any conclusion, meanwhile is very useful for qualitative analysis.

To estimate the anisotropy of particle itineracy we use the Fermi-surface-averaged squared-velocity tensor
\begin{equation}\label{eq:tensor}
    \mathcal{V}_{ij} =
    \frac{\oint_\text{FS} \frac{v_i(\mathbf{q}) v_j(\mathbf{q})} {|\mathbf{v}(\mathbf{q})|} dl}
    {\oint_\text{FS} \frac{dl} {|\mathbf{v}(\mathbf{q})|}}
\end{equation}
where the indices $i=x,y$ and $j=x,y$;
the integrals are taken over the Fermi surface (Fermi contour); 
$\mathbf{v}(\mathbf{q}) = (v_x(\mathbf{q}), v_y(\mathbf{q}))$ is a group-velocity vector; 
and $\mathbf{q}$ is a reciprocal vector that is expressed in cartesian basis in contrast to $\mathbf{k}$ that is expressed in the reciprocal-lattice basis. 
Components of $\mathbf{q}$ have units of $a_\text{hex}^{-1}$, where $a_\text{hex}$ is a hexagonal lattice parameter. The tensor components have units of $(Da_\text{hex})^2$ and this is the only quantity in our study, where the lattice parameter $a_\text{hex}$ is involved. 
The group velocities $v_i(\mathbf{q}) = \frac{\partial\omega_{m_\text{FS}}(\mathbf{q})}{\partial q_i}$ ($m_\text{FS}$ is the index of a band where the Fermi level is located) are calculated from the band structures using the centered finite-difference method and the integrals are calculated using the trapezoidal rule. The convergence with respect step sizes is ensured.
We denote the principal values (eigenvalues) of the tensor (\ref{eq:tensor}) by $\mathcal{V}_x$ and $\mathcal{V}_y$.

\begin{figure*}[t!]
    \centering
    \includegraphics[width=1\linewidth]{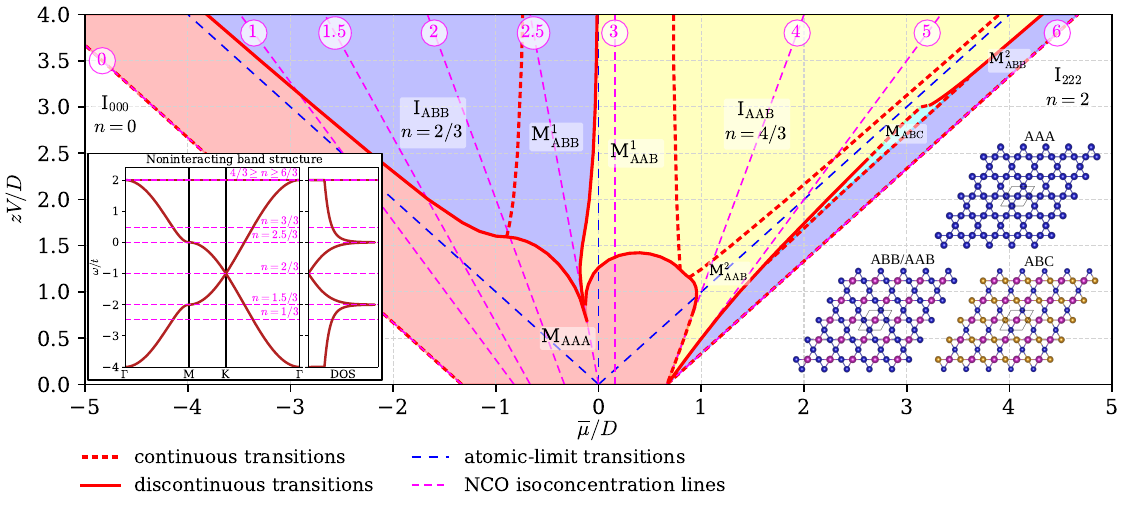}
    \caption{MFA kagome-lattice phase diagram  {with charge orders that are commensurate with the three-sublattice cell} (only phases with the lowest grand potential; no coexistence regions shown). 
    Phase naming: M is a metallic phase, I is an insulating phase; subscripts and each colored region correspond to the AAA, ABB ($n_\text{A}>n_\text{B}$), AAB ($n_\text{A}>n_\text{B}$), or ABC charge order shown on the right inset. 
    The black solid lines on the right inset indicate primitive cells of the structures; note that they are mirrored with respect to the actually used cells in calculations.
    Left inset: Kagome noninteracting band structure with shown levels that correspond to a specified total charge concentration. 
    The corresponding isoconcentration lines of a non-charge-ordered (NCO) phase are marked on a phase diagram and labeled by the value of $3n$  (number of particles per the unit cell) indicated in the circles.}
    \label{fig:phase-diagram}
\end{figure*}

\subsection{Atomic limit}\label{sec:AL}

The atomic-limit (AL) model is the model (\ref{eq:model}) with $t=0$. 
It is exactly solvable as far as the periodicity of a chosen (super)cell is enforced (in this work, the primitive cell with three lattice sites) \cite{alekseevPRB2026, AlekseevPRB2025, KapciaJSNM2019, Kapcia2021, KapciaJMMM2022}.
Despite its simplicity, it provides a limiting case ($t \rightarrow 0$ or $t \ll V, \mu$) and a useful outline and expectations for the phase diagram for both the MFA and DMFT with $U \ne 0$ \cite{alekseevPRB2026, AlekseevPRB2025}  {(cf. also \cite{Kapcia2017, AlekseevPhysicaA2026, MicnasPRB1984, ManciniPRE2008, KapciaPhysaA2016, KapciaPRE2017} for bipartite lattices)}.

For the extended Hubbard model with the NN intersite interaction (which is more general than our model, particularly, the model (\ref{eq:model}) plus a nonzero Hubbard interaction), the AL results for the kagome lattice are exactly the same as for a triangular lattice with a $\sqrt{3}\times\sqrt{3}$ supercell \cite{alekseevPRB2026, AlekseevPRB2025, Kapcia2021} (rigorous results for the ground state). 
The mapping is also valid at any temperature for the mean-field AL results.
It follows from the fact that both lattices have three sublattices and each sublattice $\alpha$ has $z/2$ NNs of the sublattice $\bar\alpha$ and $z/2$ NNs of the sublattice $\bar{\bar\alpha}$.

In the AL at the ground state, lattice sites can have integer occupations only and each phase can be named by three integers that correspond to the sublattice occupations $n_\alpha$. 
Without Hubbard interaction there are four unique phases only: $000$, $200$ (degenerate to $020$ and $002$), $220$ (degenerate to $202$ and $022$), and $222$. 
Transition lines between them are $\bar\mu = zV$ ($000$ to $200$), $\bar\mu=0$ ($200$ to $220$), and $\bar\mu = -zV$ ($220$ to $222$). 

More accurately, phases $100$ and $221$ (phases with $6$-fold degeneracy, taking into account the spin index) also have the lowest grand potential exactly on the lines $\bar\mu = \pm zV$ together with phases $000$, $200$ and $200$, $220$, respectively (phases have equal grand potentials on these lines).

Since $D=3t=0$ in the atomic limit, $zV$ and $\bar\mu$ can be expressed in arbitrary but same units which gives a good outline for a non-AL phase diagram where they are expressed in a noninteracting half-bandwidth $D$ \cite{alekseevPRB2026}.

\section{Results}\label{sec:results}

The MFA ground-state phase diagram of the kagome-lattice $t$-$V$ model with charge orders that are commensurate with the primitive cell is shown in Fig. \ref{fig:phase-diagram} for an all-encompassing range of the chemical potentials. 
Additionally, we show the AL phase diagram (i.e. just three lines: $\bar\mu=0$ and $\bar\mu=\pm zV$).

\subsection{Phase names}\label{sec:phases}

Each phase is labeled as metallic (M) or insulating (I). 
For an insulating phase, all three $A_\alpha(i\eta)$ are negligibly small, the Fermi level is located in a band gap, and the value of $3n$ is exactly integer, 
specifically, $0$ for I$_{000}$, $2$ for I$_\text{ABB}$, $4$ for I$_\text{AAB}$, and $6$ for I$_{222}$. 
Otherwise, the phase is metallic. 

The phase subscript serves to distinguish between different charge orders, as shown on the right inset in Fig. \ref{fig:phase-diagram}. 
The AAA means $n_1=n_2=n_3=n_\text{A}$ and the absence of the charge order, i.e., denotes a non-charge-ordered (NCO) phase. 
The indices $000$ and $222$ are particular cases with $n_\text{A}=0$ (fully unoccupied system) and $n_\text{A}=2$ (fully occupied system).

The AAB ($n_1=n_2=n_\text{A}$, $n_3=n_\text{B}$) and ABB ($n_1=n_\text{A}$, $n_2=n_3=n_\text{B}$) denote charge-ordered (CO) phases with the same charge order, while being different phases with discontinuous transition between them. 
In both cases, $n_\text{A} > n_\text{B}$, and thus, an AAB phase has two charge-rich and one charge-poor sublattice, whereas an ABB phase has one charge-rich and two charge-poor sublattices (similar phase distinction is found for the triangular lattice \cite{AlekseevPRB2025}). 
These are phases with $3$-fold degeneracy, e.g., the AAB is degenerate with the ABA and BAA.

The only ABC phase is a CO phase with the lowest symmetry in which all three sublattices have different occupation numbers. The phase has $6$-fold degeneracy.

Note, for convenience, each subscript is color-coded in the phase diagram (Fig. \ref{fig:phase-diagram}): AAA is red, AAB is yellow, ABB is blue, ABC is cyan, and both 000 and 222 are white.

The phase superscript is used to distinguish phases that would have the same name otherwise.

\begin{figure*}
    \centering
    \includegraphics[width=0.99\linewidth]{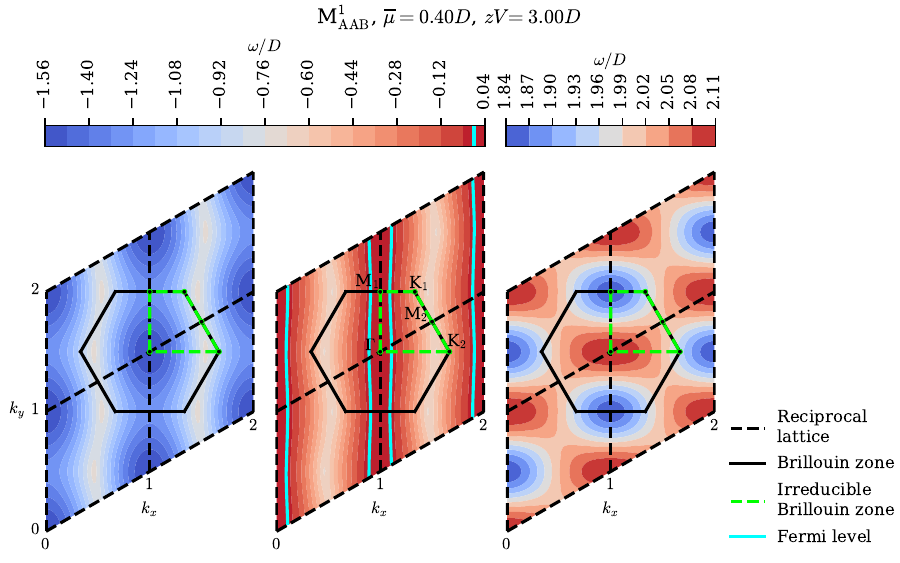}
    \caption{Band structure for the M$_\text{AAB}^1$ phase. 
    Note, the entire band-structure picture can be rotated by $60^\circ$ when choosing orders ABA or BAA, making it more similar to that shown in Fig. \ref{fig:contour-ABB}.
    The same band structure is presented in Fig. \ref{fig:band-structures}a.}
    \label{fig:contour-AAB}
\end{figure*}

\subsection{Symmetries and band structures}\label{sec:sym-bands}

The symmetry of an NCO phase (M$_\text{AAA}$, I$_\text{000}$, or I$_\text{222}$) is described by the wallpaper group $p6m$ (hexagonal Bravais lattice), with $12$ symmetry operations.
Within the MFA, its band structure is identical to that of the noninteracting kagome lattice (see the left inset in Fig. \ref{fig:phase-diagram}), except  {that} the Fermi level is shifted by $(zVn - \mu)$, i.e. by the value of $\epsilon_\text{A} = \epsilon_1 = \epsilon_2 = \epsilon_3$, see Eqs. (\ref{eq:energies}). 
It has the same structure as that of the honeycomb lattice with an additional dispersionless band on the edge; see Sec. \ref{sec:isoconcenration} for its more detailed discussion.

\begin{figure*}
    \centering
    \includegraphics[width=0.99\linewidth]{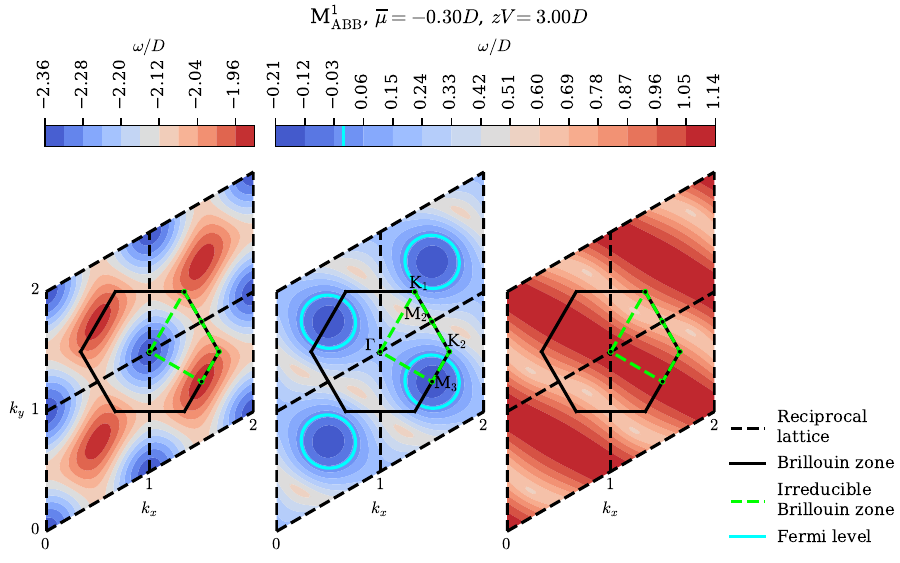}
    \caption{Band structure for the M$_\text{ABB}^1$ phase. 
    Note, the entire band-structure picture can be rotated by $60^\circ$ when choosing orders BAB or BBA, making it more similar to that shown in Fig. \ref{fig:contour-AAB}.
    The same band structure is presented in Fig. \ref{fig:band-structures}b.}
    \label{fig:contour-ABB}
\end{figure*}

AAB and ABB phases have the wallpaper group $cmm$ symmetry (centered rectangular Bravais lattice, even though the primitive cell is hexagonal, see the right inset in Fig. \ref{fig:phase-diagram}), with $4$ symmetry operations in its point group.
The irreducible Brillouin zone is four times smaller than the Brillouin zone, which is shown in Figs. \ref{fig:contour-AAB} and \ref{fig:contour-ABB} along with the definition of high-symmetry points. 
The representative band structures are shown fully in Figs. \ref{fig:contour-AAB} and \ref{fig:contour-ABB}, and shown along paths that go through the high-symmetry points in Figs. \ref{fig:band-structures}a, \ref{fig:band-structures}b, and \ref{fig:band-structures}d together with the spectral functions.

One should note that swapping the sublattice occupations (e.g., from AAB to ABA or BAA) without actually changing the system, rotates the entire band structure in the reciprocal space by $60^\circ$.

\begin{figure*}
    \centering
    \includegraphics[width=0.99\linewidth]{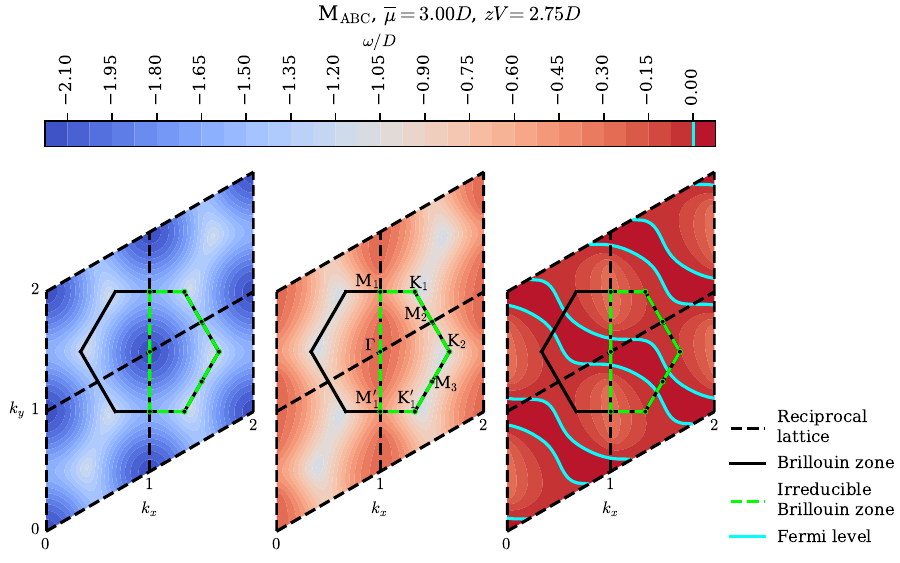}
    \caption{Band structure for the M$_\text{ABC}$ phase. 
    The same band structure is shown in Fig. \ref{fig:band-structures}c.}
    \label{fig:contour-ABC}
\end{figure*}

\begin{figure*}
    \centering
    \includegraphics[width=1\linewidth]{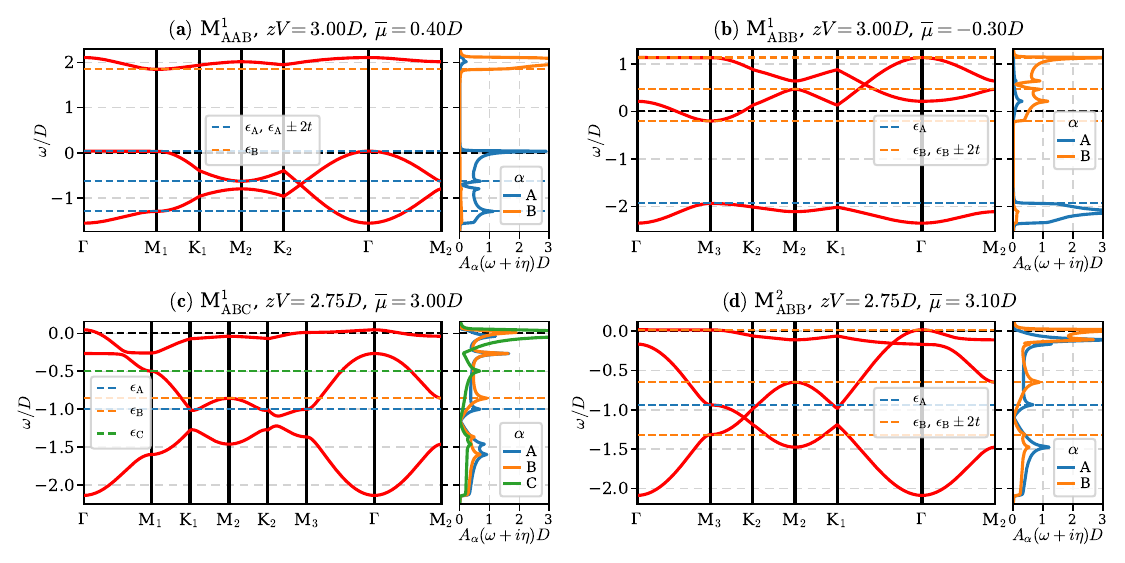}
    \caption{Band structures and sublattice-specific spectral functions of the
    (a) M$_\text{AAB}^1$ phase (same as in Fig. \ref{fig:contour-AAB}),
    (b) M$_\text{ABB}^1$ phase (same as in Fig. \ref{fig:contour-ABB}),
    (c) M$_\text{ABC}^1$ phase (same as in Fig. \ref{fig:contour-ABC}),
    and (d) M$_\text{ABB}^2$ phase together with the corresponding sublattice-specific spectral functions. 
    The horizontal dashed black lines mark the Fermi level ($\omega=0$).
    }
    \label{fig:band-structures}
\end{figure*}

The ABC phase has the wallpaper group $p2$ symmetry (oblique Bravais lattice, even though the primitive cell is hexagonal), with only $2$ symmetry operations (identity and inversion). Its irreducible Brillouin zone is two times smaller than the Brillouin zone, as shown in Fig. \ref{fig:contour-ABC} along with the definition of the high-symmetry points. 
It also shows a representative band structure, while the band structure along a path in the irreducible Brillouin zone is shown in Fig. \ref{fig:band-structures}c together with the spectral functions.

Due to the inversion symmetry that is present for any charge order, all points K are symmetrically-equivalent (which is not true about the lines between them and other high-symmetry points).

The evolution of the order parameters and band structures for $zV=2.75D$ from $\bar\mu = 2.5D$ to $3.5D$ (across continuous transitions I$_\text{AAB}$--M$_\text{AAB}^2$--M$_\text{ABC}$--M$_\text{ABB}^2$--I$_{222}$), is shown in the Supplemental Material file \texttt{V2.75.mp4}. It also shows charge polarizations  {$\Delta_{12} = (n_1 - n_2)/2$ and $\Delta_{23} = (n_2 - n_3)/2$}.

In the large-$V$ limit, the band structures of the I$_\text{AAB}$ and M$_\text{AAB}^1$ phases become very similar to a band structure of a noninteracting 1D chain formed by lattice sites of the sublattice A (see the right inset in Fig. \ref{fig:phase-diagram}), which is well-separated from an atomic-like (nearly dispersionless) band that comes from the sublattice B.  {As} $V$ decreases, the hybridization between A and B sublattices increases, the lower band increasingly deviates from the 1D-chain band structure and the upper ``level'' broadens into a band. 

The same applies to the I$_\text{ABB}$ and M$_\text{ABB}^1$ phases except  {that} the bands are swapped: The upper band is the 1D-like band of the sublattice B and the lower band is an atomic-like one of the sublattice A.

As found for the triangular lattice \cite{AlekseevPRB2025}, it is useful to follow the levels $\epsilon_\alpha$ in the band-structure and spectral-function plots. 
Moreover, similarly to features of a honeycomb lattice, which serves as a sublattice within the $\sqrt{3}\times\sqrt{3}$ triangular-lattice supercell, here, the 1D-lattice features are found in the analysis of band structures. 
The 1D-lattice density of states is symmetric and has two features only: singularities on edges of the band structure at $\pm 2t$. 
Consequently, besides $\epsilon_\alpha$, we look at $\epsilon_A \pm 2t$ for AAB phases (sublattice A forms a 1D lattice) and $\epsilon_B \pm 2t$ for ABB phases (sublattice B forms a 1D lattice).

For the phases I$_\text{AAB}$, M$_\text{AAB}^1$, and M$_\text{AAB}^2$ the $\epsilon_B$ is always fixed to the bottom of the narrow upper band (see Fig. \ref{fig:band-structures}a) at the point M$_1$; the $\epsilon_A$ is fixed to the valley in the middle of the lower band at the point M$_2$; the $\epsilon_A - 2t$ is fixed to the saddle point at the point M$_1$; and the $\epsilon_A + 2t$ is fixed to the top of the lower band in a line $\Gamma$--M$_1$. 
Thus, the band gap equals exactly $\epsilon_B - \epsilon_A - 2t = \frac{zV}{2} (n_\text{A} - n_\text{B}) - 2t$. Similar arrangement is found for the I$_\text{ABB}$ and M$_\text{ABB}^1$ phases as shown in Fig. \ref{fig:band-structures}b, with the band gap $\frac{zV}{2} (n_\text{B} - n_\text{A}) - 2t$.

Importantly (see Sec. \ref{sec:quasi-1D}), both in AAB and ABB phases, the shape of the top of the 1D-like bands ($\epsilon_\alpha + 2t$) stays nearly unchanged for various $V$ or various degree of hybridization between the A and B sublattices. 
In contrast, the bottom of these bands (around $\epsilon_\alpha - 2t$) changes its shape: Compare, e.g., the $\Gamma$--M$_1$ line in Fig. \ref{fig:band-structures}a with the $\Gamma$--M$_3$ line in Fig. \ref{fig:band-structures}b, both of which become dispersionless in the large-$V$ limit.

The band gap in the I$_\text{AAB}$, M$_\text{AAB}^1$, M$_\text{AAB}^2$, I$_\text{ABB}$, and M$_\text{ABB}^1$ phases disappears closer to the transition to an NCO phase, the points $\epsilon_\alpha$ and $\epsilon_\alpha \pm 2t$ become increasingly mixed, but they are always fixed to extremum or saddle points (band edges or singularities in the spectral function). 
One example is the M$_\text{ABB}^2$ phase shown in Fig. \ref{fig:band-structures}d, which is very close to the noninteracting kagome-lattice band structure (close to the continuous transition to the I$_{222}$ phase), but $\epsilon_A$, $\epsilon_B$, and $\epsilon_B \pm 2t$ always correspond to singularities in the spectral function. 

Similarly, energy levels $\epsilon_\alpha$ always correspond to specific singularities (extrema and saddle points of band structures) for the M$_\text{ABC}$ phase, as shown in Fig. \ref{fig:band-structures}c.

\subsection{Stripes and quasi-1D structure}\label{sec:quasi-1D}

In AAB and ABB phases, the sublattices A and B, respectively, form a 1D-stripe structure (see the right inset in Fig. \ref{fig:phase-diagram}). 
In the M$_\text{AAB}^1$ phase, the Fermi level is located at the top of the lower band (Figs. \ref{fig:contour-AAB} and \ref{fig:band-structures}a). 
This band is formed primarily by the sublattice A (see sublattice-specific spectral functions). 
Meanwhile, the hybridization between the sublattices A and B contributes to the regions of the lower band that are away from its top. In other words, only the sublattice A contributes to the spectral weight at the Fermi level ($A_\text{B}(i\eta) \approx 0$), the charge in the sublattice B is localized, the sublattice-A stripes are isolated from the sublattice B, and the structure is quasi-1D; that is, charge itineracy occurs only in one direction. 
One can see in Fig. \ref{fig:contour-AAB} that the Fermi level is in the band that is dispersionless in one direction.
Emphasizing the particle-hole asymmetry, it is not the case for the M$_\text{ABB}^1$ phase, where the Fermi level is located at the bottom of similar 1D-like bands, which is more affected by the hybridization between A and B sublattices (Figs. \ref{fig:contour-ABB} and \ref{fig:band-structures}b).

This phenomenon is  {analogous to} formation of a pinball-liquid phase in the triangular lattice \cite{hotta2006, hotta2007, miyazaki2009variational, trousselet2012valence, Ralko2015, AlekseevPRB2025, alekseevPRB2026}, in which the honeycomb lattice is conducting, while the charge in the third sublattice is localized.

The quasi-1D phase disappears close to the NCO phase (M$_\text{AAA}$), but the exact point of this very smooth phase transition (order parameter $A_\text{B}(i\eta)$ asymptotically becomes finite) depends on an arbitrarily chosen threshold for the $A_\text{B}(i\eta)$, and therefore we do not show it.

The unusual M$_\text{ABB}^2$ phase also shows quasi-1D behavior when it is not close to the I$_{222}$ phase (see Supplemental Material file \texttt{V2.75.mp4}) despite the strong hybridization between sublattices A and B, which does not contribute to the spectral weight at the Fermi level (Fig. \ref{fig:band-structures}d). 
There is no counterpart of such a phase for the triangular lattice. 
Interestingly, the band structure of the M$_\text{ABB}^2$ phase is more similar to that of the M$_\text{AAB}^2$ phase, instead of the M$_\text{ABB}^1$ phase (see Supplemental Material file \texttt{V2.75.mp4}). 
At the same time, judging by locations of levels $\epsilon_\text{A}$, $\epsilon_\text{B}$, and $\epsilon_\text{B} \pm 2t$, it can be viewed as the M$_\text{ABB}^1$ phase where the Fermi level is exactly where it needs to be to form the quasi-1D structure: at the top of the bands that have previously been the 1D-like bands, $\epsilon_\text{B} + 2t$ (Figs. \ref{fig:band-structures}b and \ref{fig:band-structures}d).

To quantify the anisotropy of itineracy and the quasi-1D behavior, Table \ref{tab:1} shows $\mathcal{V}_x$ and $\mathcal{V}_y$ of representative ABB and AAB phases. The $\mathcal{V}_x$ and $\mathcal{V}_y$ are the eigenvalues of $\mathcal{V}_{ij}$, i.e., its components along the principal axes (eigenvectors). For convenience, instead of the ABB order for the ABB phases, we use the BBA arrangement which rotates the band structure in Fig. \ref{fig:contour-ABB} by $60^\circ$ clockwise. Due to this, the principal axes of all phases in Table \ref{tab:1} have the same directions. Particularly, the principal direction $x$ is horizontal to Fig. \ref{fig:contour-AAB} and the principal direction $y$ is vertical to Fig. \ref{fig:contour-AAB}.
The quasi-1D behavior is evident from the orders of difference between $\mathcal{V}_x$ and $\mathcal{V}_y$ for the M$_\text{AAB}^1$ and M$_\text{BBA}^2$ phases.

\begin{table}[]
    \caption{\label{tab:1}%
    The components of the Fermi-surface-averaged squared-velocity tensor (eq. (\ref{eq:tensor})) along the principal axes
    for representative phases with the stripe charge orders
    (up to three significant digits).}
    \centering
    \begin{tabular}{|c|c|c|}
        \hline
         {Phase} & $\mathcal{V}_x / (Da_\text{hex})^2$ & $\mathcal{V}_y / (Da_\text{hex})^2$
        \\ \hline
        \shortstack{M$_\text{BBA}^1$ \\ $zV=3.00D$, $\bar\mu=-0.3D$} & $0.948$ & $0.773$
        \\ \hline
        \shortstack{M$_\text{AAB}^1$ \\ $zV=3.00D$, $\bar\mu=0.4D$} & $0.569$ & $5.16 \cdot 10^{-4}$
        \\ \hline
        \shortstack{M$_\text{AAB}^2$ \\ $zV=2.75D$, $\bar\mu=2.8D$} & $0.0949$ & $0.220$
        \\ \hline
        \shortstack{M$_\text{BBA}^2$ \\ $zV=2.75D$, $\bar\mu=3.1D$} & $0.101$ & $1.53 \cdot 10^{-3}$
        \\ \hline
    \end{tabular}   
\end{table}

\subsection{Atomic-limit prediction}

There is partial correspondence between the AL and MFA phase diagrams. 
The phases I$_{000}$, M$_\text{AAA}$, and I$_{222}$ are counterparts of the NCO AL phases $000$, $111$, and $222$; the phases I$_\text{ABB}$ and M$_\text{ABB}^1$ are counterparts of the AL phase $200$; and the phases I$_\text{AAB}$ and M$_\text{AAB}^1$ are counterparts of the AL phase $220$. 
It is particularly evident in the large-$V$ limit. 

Nevertheless, the finite charge itineracy on the kagome lattice gives rise to unexpected phases for $\bar\mu>0$, between the AL line ($\bar\mu = zV$) and the fully occupied phase ($\bar\mu = zV + 2t$, see Sec. \ref{sec:isoconcenration}). 
These phases are M$_\text{AAB}^2$, M$_\text{ABB}^2$, and the small-range intermediate phase M$_\text{ABC}$. 
In the triangular lattice, the likely reason for similar behavior in the same region is a Van Hove singularity \cite{AlekseevPRB2025, alekseevPRB2026} of a noninteracting system, but this gives rise to one unexpected phase only. Here, the noninteracting system has a Dirac-delta level at the very edge, which impacts much more. Its role is discussed in greater detail in Sec. \ref{sec:isoconcenration}.

Similarly to the triangular lattice, this unexpected region involves a continuous transition from a CO phase directly to the fully occupied NCO phase. 
This makes the I$_{222}$ phase more important than it usually is in lattice models. This transition holds for $U \ne 0$ within the DMFT \cite{alekseevPRB2026}.

\subsection{Phase transitions}\label{sec:transitions}

The continuous, or second-order, transitions are those where the order parameters ($n_\alpha$ and $A_\alpha(i\eta)$) change continuously. 
The simplest cases of these transitions are transitions I$_\text{ABB}$--M$_\text{ABB}^1$, M$_\text{AAB}^1$--I$_\text{AAB}$--M$_\text{AAB}^2$, and I$_{000}$--M$_\text{AAA}$--I$_{222}$. 
They can be characterized as smooth movement of the Fermi level between the bands; see Supplemental Material file \texttt{V2.75.mp4} for the I$_\text{AAB}$--M$_\text{AAB}^2$ transition. 

The discontinuous, or first-order, transitions are those that involve a jump of the order parameters, as well as their hysteresis or the coexistence regions of two phases. 
The transition point is defined by the equality of the grand potentials $\Omega$ of two phases, i.e., during transition, $\Omega$ changes  {continuously}, but its derivative (with respect to a model parameter) on both sides is different.

A discontinuous transition is common between the colored regions of the phase diagram (Fig. \ref{fig:phase-diagram}), such as the ABB (blue) to AAA (red) region, the AAB (yellow) to AAA (red) region, or the ABB (blue) to AAB (yellow) region. 
The M$_\text{ABC}$ phase can be viewed as an intermediate phase through which the continuous transition between the yellow M$_\text{AAB}^2$ phase and blue M$_\text{ABB}^2$ phase is possible.

\begin{figure}
    \centering
    \includegraphics[width=1\linewidth]{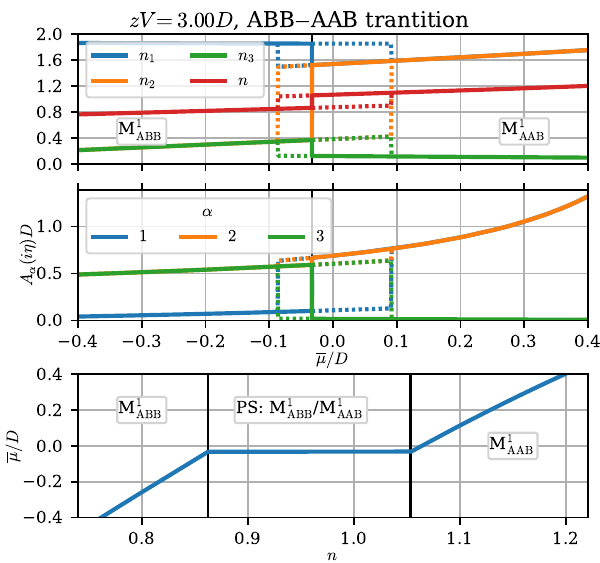}
    \caption{The discontinuous transition between the ABB and ABB regions around $\bar\mu = 0D$ for $zV = 3D$. The dotted lines represent hysteresis, i.e., the parameters of metastable (or, possibly, unstable) phases. The vertical black lines show the phase transitions. The phase-separated (PS) state is also shown.}
    \label{fig:hysteresis}
\end{figure}

A discontinuous transition implies the existence of phase-separated states. 
In Fig. \ref{fig:hysteresis} we show the hysteresis between the M$_\text{ABB}^1$ and M$_\text{AAB}^1$ phases, and show the phase-separation region in the lower panel of the Figure. For $zV=3D$, the total densities in the range $0.862 < n < 1.054$, including the half filling, can be found within the phase-separated states only. Within the fixed-occupation approach, to find the phase separation, one would require the Maxwell construction but it is naturally taken into account in the grand-canonical ensemble with $\mu$ as a variable \cite{BakAPPA2004}. 
Note that the phases with higher ground potential (dotted lines in Fig. \ref{fig:hysteresis}) are simply those for which the self-consistent solution of the model (\ref{eq:MFmodel}) has been found, but the phases may not only be metastable but also unstable.

We have found that, similarly to the triangular lattice \cite{AlekseevPRB2025}, a discontinuous transition takes place when the Fermi level approaches a band edge or a singularity in a spectral function. For example, for the M$_\text{AAB}^1$ phase (Fig. \ref{fig:band-structures}a), the band gap decreases and disappears when $V$ decreases, and it is destabilized toward  {the} NCO M$_\text{AAA}$ phase when the Fermi level ($\omega = 0$) nearly meets the upper band ($\omega = \epsilon_\text{B}$, see Fig. \ref{fig:band-structures}a).

However, there are parameter ranges in which the Fermi level of a CO phase does not meet a compromising level (band edge or singularity) until the very phase transition to the NCO phase, ultimately making the phase transition continuous and creating tricritical points in the phase diagram. This way, both the symmetry breakings (transitions from NCO to CO phases) M$_\text{AAA}$--M$_\text{AAB}^1$ and M$_\text{AAA}$--M$_\text{ABB}^1$ are continuous at $\bar\mu = -zV / 6$; the symmetry breaking M$_\text{AAA}$--M$_\text{AAB}^2$ is continuous along the line $\bar\mu = zV / 3 + 2t$; and the symmetry breaking I$_{222}$--M$_\text{ABB}^2$ is continuous along the line $\bar\mu = zV + 2t$. 
These lines are discussed in Sec. \ref{sec:isoconcenration}. Such  {behavior} can likely be found within the DMFT with $U \ne 0$, where the band-structure analysis is no longer available \cite{alekseevPRB2026}.

Note, self-consistent approaches with either numerical integration or summation over the $\mathbf{k}$ points do not allow us to find the exact location of tricritical points, such as along the lines $\bar\mu = -zV / 6$ or $\bar\mu = zV / 3 + 2t$ due to their extremely asymptotic behavior \cite{AlekseevPhysicaA2026}. 
This has been found for the simplest case of the Bethe lattice for which the analytical simplifications are available and make identification of such points possible \cite{AlekseevPhysicaA2026}. 
It is reasonable to assume that both M$_\text{AAA}$--M$_\text{AAB}^1$ and M$_\text{AAA}$--M$_\text{ABB}^1$ continuous symmetry-breaking transitions take place at exactly $zV = +0$ ($\bar\mu = -zV / 6 = 0$, see Fig. \ref{fig:phase-diagram}), but our numerical method fails to resolve such small deviations between two phases even for an extremely dense $\mathbf{k}$-point grid. 

Since the exact identification of tricritical points potentially impossible within the approach that is used in our research, when we plot the Fig. \ref{fig:phase-diagram} we use the following criteria: The transition is continuous where the hysteresis between two phases is not found along the lines with $0.001D$ step for model parameters $zV$ and $\bar\mu$.

\subsection{NCO isoconcentration lines}\label{sec:isoconcenration}

For the noninteracting (tight-binding) kagome-lattice model and with the chosen sign convention for the hopping term, the Fermi level ($\omega=0$) is located at the upper Van Hove singularity; the lower Van Hove singularity is at $\omega=-2t$; the Dirac cone is at $\omega = -t$, the lower edge is at $\omega = -4t$; the upper edge and the Dirac-delta peak are at $\omega = 2t$ \cite{hanisch1997}. 

In the MFA, within the NCO phases (ultimately, the M$_\text{AAA}$ phase) the total charge density is the same along straight lines defined as
\begin{equation}
    \bar\mu = zV (n - 1) + \mu_0(n)
\end{equation}
(NCO isoconcentration lines), where $\mu_0(n)$ is a chemical potential to achieve the total density $n$ in the noninteracting ($V=0$) model.

These lines are of interest for several values of the total density, particularly for the densities that correspond to the noninteracting features (Van Hove singularities, Dirac-delta, Dirac-cone, band edges), and for the densities with an integer value of $3n$. They are shown in the phase diagram (Fig. \ref{fig:phase-diagram}), and the corresponding levels are shown in the left inset of Fig. \ref{fig:phase-diagram}. 
Let us discuss each line:
\begin{itemize}
    \item{$3n = 0$ and $\mu_0(0) = -4t$. The line of the continuous transition to the fully unoccupied phase. Usefully, its position is exactly defined and it stays this way within the DMFT for $U \ne 0$ \cite{alekseevPRB2026}.}
    \item{$3n = 1$ and $\mu_0(1/3) = -2.47148t$. We expect site-selective Mott phases along this line within the DMFT for $U \ne 0$ (the corresponding AL phase is $100$).}
    \item{$3n = 1.5$ and $\mu_0(1/2) = -2t$. Despite the Van Hove singularity in the NCO phase, we do not see distinctive features in the phase diagram along this line, at least for $U=0$.  {This contrasts with the behavior at the other} Van Hove singularity at $3n =  2.5$.}
    \item{$3n = 2$ and $\mu_0(2/3) = -t$. The line of the Dirac cone in the NCO phase and the line along which the band insulator (I$_\text{ABB}$) emerges  {as $V$ increases} (the corresponding AL phase is $200$). 
    Along this line we also expect site-selective Mott phases within the DMFT for $U \ne 0$ (the corresponding AL phase is $110$).}
    \item{$3n = 2.5$ and $\mu_0(5/6) = 0t$. The line of the Van Hove singularity. 
     It is exactly the line of the tricritical point and the asymptotic continuous symmetry breaking from the NCO phase to two CO phases at the same time, see Sec. \ref{sec:transitions}.}
    \item{$3n = 3$ and $\mu_0(1) = 0.47150t$. We expect site-selective Mott phases along this line within the DMFT for $U \ne 0$ (the corresponding AL phases are $111$ (NCO Mott insulator) and $210$). 
    It is the vertical line in the MFA phase diagram, as imposed by our definition of the $\bar\mu$ ($\bar\mu = \mu - zV$).}
    \item{$3n = 4$ and $\mu_0(4/3) = 2t$. All occupations from $3n = 4$ to $3n = 6$ originate from the same $V=0$ point, due to the Dirac-delta peak (dispersionless band). 
    As a result, the entire region between the line $\bar\mu = zV / 3 + 2t$ and the line $\bar\mu = zV + 2t$ in the phase diagram is rather anomalous. 
    The $3n = 4$ NCO isoconcentration line exactly corresponds to the continuous phase transition from M$_\text{AAA}$ to M$_\text{AAB}^2$. 
    It is also the line along which the band insulator (I$_\text{AAB}$) emerges as $V$ increases (the corresponding AL phase is $220$), and the line along which we expect site-selective Mott phases within the DMFT for $U \ne 0$ (the corresponding AL phase is $211$).
    Note, this line creates an unusual reentrance to an NCO phase (from I$_{222}$ to M$_\text{AAA}$ through the CO phases) as the intersite interaction $V$ increases.}
    \item{$3n = 5$ and $\mu_0(5/3) = 2t$. Exactly the line of the discontinuous transition from M$_\text{AAB}^2$ to M$_\text{ABB}^2$, and the line along which we expect site-selective Mott phases within the DMFT for $U \ne 0$ (the corresponding AL phase is $221$).}
    \item{$3n = 6$ and $\mu_0(2) = 2t$. The line of the continuous transition to the fully occupied phase. Usefully, its position is exactly defined and it stays this way within the DMFT for $U \ne 0$ \cite{alekseevPRB2026}.}
\end{itemize}

Within the DMFT but with the intersite-interaction term treated on the Hartree level, the $\mu_0(n)$ is also a nonanalytic function of $U$, but the NCO isoconcentration lines as a function of $V$ are still the straight lines \cite{alekseevPRB2026}.
Note, within the DMFT with $U \ne 0$, more site-selective Mott phases can be expected when larger supercells  {are used}, which we disregard in this research \cite{alekseevPRB2026, Kapcia2017}.

\section{Summary}\label{sec:summary}

Even the mean-field phase diagram of the kagome lattice exhibits a variety of unusual phenomena. 
Similarly to the triangular lattice \cite{AlekseevPRB2025}, it has strong particle-hole asymmetry; phase transitions whose order changes from first to second; the continuous phase transition from a CO phase directly to the fully occupied NCO phase. Additionally, for positive $\bar\mu$ (electron doping) the quasi-1D structures form in which a channel for the charge itineracy opens in one direction only, which is fully analogous to the triangular-lattice pinball-liquid phase. 

Besides that, the existence of the Dirac-delta peak in the noninteracting kagome lattice gives rise to anomalous CO phases between the lines $\bar\mu=zV$ (AL phase transition between phases $220$ and $222$) and $\bar\mu = zV + 2t$ (exactly defined transition to the fully occupied phase), including the phases M$_\text{ABC}$ and M$_\text{ABB}^2$ that do not have counterparts in the triangular lattice.

The kagome lattice features the CO phases even for arbitrary small nonzero intersite interaction $V$, but we may expect that this will be suppressed by the finite onsite interaction, similarly to the Bethe lattice (MFA in Ref. \cite{AlekseevPhysicaA2026}, DMFT in Ref. \cite{Kapcia2017}).

The presented results are for the spin-$1/2$ $t$-$V$ model, which we view as a $U=0$ point of the extended Hubbard model. 
The $U \neq 0$ case should be treated within the DMFT to provide a more accurate description of onsite correlations.
This approach is much more computationally expensive and does not have access to band-structure analysis, while meets even more serious convergence problems. 
However, it has been is found that the MFA results contain features that are still present in the DMFT \cite{alekseevPRB2026}, where their analysis is harder, such as varying order of phase transitions, the phases with weak onsite correlations (band insulators and nearby metallic phases), or quasi-1D structures. 
By combining our results with the AL results, certain predictions about the DMFT phase diagram have been made. We expect our results to be equally useful in the extended DMFT studies that take the nonlocal correlations into account, which is a very important step in the study of frustrated-lattice 2D materials. 

The scope of this study is limited to the charge orders that are commensurate with the primitive cell, but longer-period orders are possible and especially relevant close to the fully occupied and fully unoccupied lattice, with further than nearest-neighbor interaction, and with $U \ne 0$. Additionally, our approach does not allow to capture the incommensurate orders which are not ruled out as well.


\section*{Declaration of competing interest}

The authors declare that they have no known competing financial interests or personal relationships that could have appeared to influence the work reported in this paper.
The authors' institution had no role in the design of the study; in the collection, analyses, or interpretation of data; in the writing of the manuscript, or in the decision to publish the results. 

\section*{Data availability}

Data will be made available on request. 
All presented data are obtained by (numerical) solving the equations, which forms are included explicitly in the present work.

\section*{Funding sources}
This research did not receive any specific grant from funding agencies in the public, commercial, or not-for-profit sectors.

\printcredits

\bibliographystyle{elsarticle-num}

\end{document}